# A computational alloy design framework for the promotion of amorphous grain boundary complexions

Prince Sharma [1,2], Jaime Marian [3,4], Jason R. Trelewicz [5,6,7], Timothy J. Rupert [1,2,*]

[1] Hopkins Extreme Materials Institute, Johns Hopkins University, Baltimore, MD 21218, USA
[2] Department of Materials Science and Engineering, Johns Hopkins University, Baltimore, MD 21218, USA
[3] Department of Materials Science and Engineering, University of California Los Angeles, Los Angeles, CA 90095, USA
[4] Department of Mechanical and Aerospace Engineering, University of California Los Angeles, Los Angeles, CA 90095, USA
[5] Department of Materials Science and Chemical Engineering, Stony Brook University, Stony Brook, NY, USA
[6] Institute for Advanced Computational Science, Stony Brook University, Stony Brook, NY, USA
[7] Materials Science and Technology Division, Oak Ridge National Laboratory, Oak Ridge, TN, USA
* Corresponding Author: tim.rupert@jhu.edu

**Abstract**:

Amorphous grain boundary complexions have been shown to be radiation tolerant interfaces that can also reduce grain boundary embrittlement, marking them as favorable microstructural features. However, the incorporation of these features into new alloy systems is often a slow and arduous process based on trial and error. Here, a computational framework for alloy design is presented which enables the selection of dopants that promote the formation of amorphous grain boundary complexions. This framework is primarily built on density functional theory calculations and is demonstrated for W-rich binary and ternary alloys, which represent a promising target for fusion energy materials. Our framework first evaluates the grain boundary segregation tendency of dopants and then the energy penalty for amorphization alongside targeted interfacial energy comparison, with the end goal of identifying the best dopants. For a W base, Y and some transition metals such as Co and Ni are found to significantly lower these energetic barriers. Electronic structure analysis, local lattice distortion, and charge density distributions are calculated and used to provide mechanistic explanations for these dopant selections. Finally, the framework is validated by comparing with experimental literature for W alloys and a refractory complex concentrated alloy, showing a strong correlation between our dopant selections and low sintering onset temperatures that have been attributed to activated sintering. As a whole, this work establishes a transferable pipeline for designing alloys with grain-boundary complexions across diverse alloy systems.

## 1. Introduction

Amorphous grain boundary complexions have emerged as a transformative class of microstructural features for engineering advanced materials.[1] Unlike conventional grain boundaries, amorphous complexions are thermodynamically-stable, structurally-disordered interfacial states that offer a unique combination of enhanced processability and superior performance in extreme environments.[2-8] Despite their promise, the discovery of alloy systems that support these complexions remains a significant challenge. Current selection rules are largely empirical, such as those established by Schuler and Rupert[9] that emphasize common predictors of metallic glasses. Similarly, while thermodynamic models from Luo and coworkers[10,11] successfully identified amorphous complexions as stable interfacial states, these models often rely on somewhat simplified bulk approximations. For example, it is common to invoke empirical Miedema-type rules whose underlying assumptions preclude a proper quantum-mechanical description of metallic bonding. Pettifor[12] demonstrated that alloy heats of formation arise primarily from changes in metallic bond order rather than charge-transfer ionic contributions, underscoring the value of first-principles calculations for interfacial penalties and amorphization energetics. While the relatively simple models are excellent for rapid screening, more accurate treatments and a robust computational pipeline are needed once a target system is chosen. In addition, prior work has primarily focused on binary alloys, while multicomponent systems are both scientifically and technologically important.

W and W-rich alloys are systems where amorphous complexions would be particularly beneficial due to their critical application in fusion energy technologies. These materials are primary candidates for plasma-facing components in fusion reactors due to the high melting point of W (3695 K), low sputtering yield under plasma particle bombardment, and resistance to neutron-induced swelling.[13-15] However, conventional processing of W often requires extremely high temperatures, which leads to excessive grain growth and degraded mechanical properties.[16-18] In addition, radiation-induced defects accumulate during reactor operation, causing embrittlement and limiting component lifetime.[2,19-22] Amorphous complexions offer a potential pathway to solve both issues.

First, amorphous complexions can enable *activated sintering*, defined as a rapid increase in sample densification while remaining in the solid state, which offers a route to process W at reduced temperatures through the addition of modest amounts of metal dopants. Hayden and Brophy first reported that Ni additions enable densification of W powders at temperatures 600-800 °C below those required for pure W.[23] German and Munir subsequently demonstrated that Pd, Fe, and Co also enhance sintering, while Cu and Mo show no beneficial effects.[24] The activated sintering mechanism involves the formation of stable, nanoscale liquid-like films at grain boundaries below the bulk solidus temperature, facilitating mass transport while maintaining structural stability.[10,11] As defined by Dillon and Harmer,[25-27] these interfacial films would be grain boundary complexions due to their constant thickness and stability under the given set of processing conditions. These structurally disordered, quasi-liquid complexions at grain boundaries enhance mass transport without macroscopic melting. Research by Luo and coworkers has explicitly studied these amorphous complexions and the conditions under which they form.[1,10,11,28-31] For W-Ni, high-resolution transmission electron microscopy revealed disordered, Ni-enriched layers of 0.6-1 nm thickness at the grain boundaries.[11,30]

Beyond processing, amorphous complexions can induce mechanical toughening and improved radiation tolerance, both key performance targets for W. Grain boundaries can act as sinks for radiation-induced point defects, yet traditional ordered GBs often exhibit a sink bias where they preferentially absorb interstitials while leaving vacancies to cluster into voids. Amorphous complexions lack the ordered atomic arrangements of traditional grain boundaries and have a broad distribution of atomic free volume levels, meaning they can absorb both vacancies and interstitials with high efficiency to facilitate defect annihilation[2,5]. Amorphous complexions can also stabilize nanocrystalline grain sizes at high temperatures, ensuring that a high density of grain boundary sinks remain available during service. Nanocrystalline metals tend to be relatively brittle,[32-34] with grain boundary fracture being a particularly problematic failure mode.[35,36] Amorphous complexions have been shown to increase malleability[7] and ductility[37] in Cu-rich nanocrystalline alloys by acting as efficient sinks for incoming dislocations[8,38], suggesting that they are generally promising features for toughening of the grain boundary region.

In this study, a predictive framework based on first-principles calculations is built for identifying dopants that promote amorphous complexion formation. Rather than relying on empirical rules, this computational pipeline calculates segregation tendencies, the energy penalty required to transition from a crystalline to an amorphous state, and the specific interfacial energies that dictate complexion stability. This framework is explored in the context of W-rich alloys, investigating various combinations of transition metal dopants in both binary and ternary systems. Dopant selection as well as optimal grain boundary concentrations are explored, with electronic structure analysis used to demonstrate why specific dopant chemistries promote interfacial disorder while others resist it. Our approach is subsequently benchmarked against experimental reports of activated sintering for both W-rich alloys and a refractory complex concentrated alloy, demonstrating a transferable methodology for discovering amorphous complexions in entirely new material systems.

## 2. Methods

A qualitative thermodynamic description for amorphous intergrainular films was first stated by Keblinski et. al.[39] Prior to this, Raj[40] had already analyzed the thermodynamic barriers to crystallization of confined intergranular glass films in $Si_3N_4$, providing the conceptual foundation for the stability of amorphous intergranular films. The key energy-balance inequality showing that a thin amorphous film can be thermodynamically-stable was stated by Wang and Chiang[41] and was later explicitly formulated and quantitatively developed by Luo and Shi,[10] providing a foundation for our computational pipeline. This model predicts that amorphous intergranular films form when the relatively high energy of a crystalline grain boundary can be reduced by replacing it with two lower energy crystal-liquid interfaces and an amorphous film. This condition is expressed as:

$$\Delta G_{amorph} \cdot h < \gamma_{GB} - 2 \cdot \gamma_{cl} \equiv \Delta\gamma \qquad (1)$$

where $\Delta G_{amorph}$ is the volumetric free energy for forming an undercooled liquid or amorphous phase below the bulk solidus temperature, $h$ is the film thickness, $\gamma_{GB}$ is the energy of a high-angle grain boundary without adsorption, and $\gamma_{cl}$ is the crystal-liquid interfacial energy. The thermodynamic tendency to stabilize a quasi-liquid intergranular film can be represented by a parameter, λ, defined as:

$$\lambda = -\Delta\gamma / \Delta G_{amorph} \qquad (2)$$

Larger λ values indicate greater thermodynamic driving force for complexion formation and scale with the equilibrium film thickness. This parameter can be computed by combining bulk thermodynamic data with interfacial energy estimates, enabling the construction of grain boundary phase diagrams that predict complexion transitions as functions of temperature and composition.[29]

**Figure 1** shows our framework for predicting amorphous complexion formation ability, with individual calculations being performed with first-principles methods (additional details are below). In the first step, we restrict ourselves to dopants that will segregate to the grain boundaries. Otherwise, one cannot achieve significant changes in local grain boundary chemistry to activate a premelting transition. The equation used to calculate grain boundary segregation energy ($\Delta E_{seg,GB}$) is:

$$\Delta E_{seg,GB} = E_{GB}^{solute} - E_{bulk}^{solute} \quad (3)$$

where $E_{GB}^{solute}$ is the energy of dopant at the grain boundary and $E_{bulk}^{solute}$ is the energy of dopant in the bulk. A negative value means that dopant atoms prefer to segregate to the grain boundary. In the second step, the relative stability of amorphous and crystalline phases is calculated, represented here as the amorphous stabilization energy ($\Delta E_{stab}$). This quantity captures the energy difference between the two competing phases at a given composition and can be determined using:

$$\Delta E_{stab} = E_{form}^{amorphous} - E_{form}^{\min(crystal)} \quad (4)$$

where $E_{form}^{amorphous}$ is the formation energy for an amorphous structure while $E_{form}^{\min(crystal)}$ is the formation energy for the most stable crystalline structure. A similar approach comparing formation energies between amorphous and crystalline phases has been proven effective to predict accelerated amorphization of high-entropy alloys.[42] Dopants that reduce $\Delta E_{stab}$ will lower the energy penalty for forming amorphous grain boundary complexions (provided they segregate to the grain boundary), thereby promoting complexion stability according to Equations (1) and (2). In the third and final step, the energy penalty associated with an interface between amorphous and crystalline regions can be calculated (details provided below) and subsequently compared with the original grain boundary energy and $\Delta E_{stab}$. Generally, one would want a segregating dopant that has a low, possible negative, amorphous stabilization energy and low amorphous-crystalline interfacial energies.

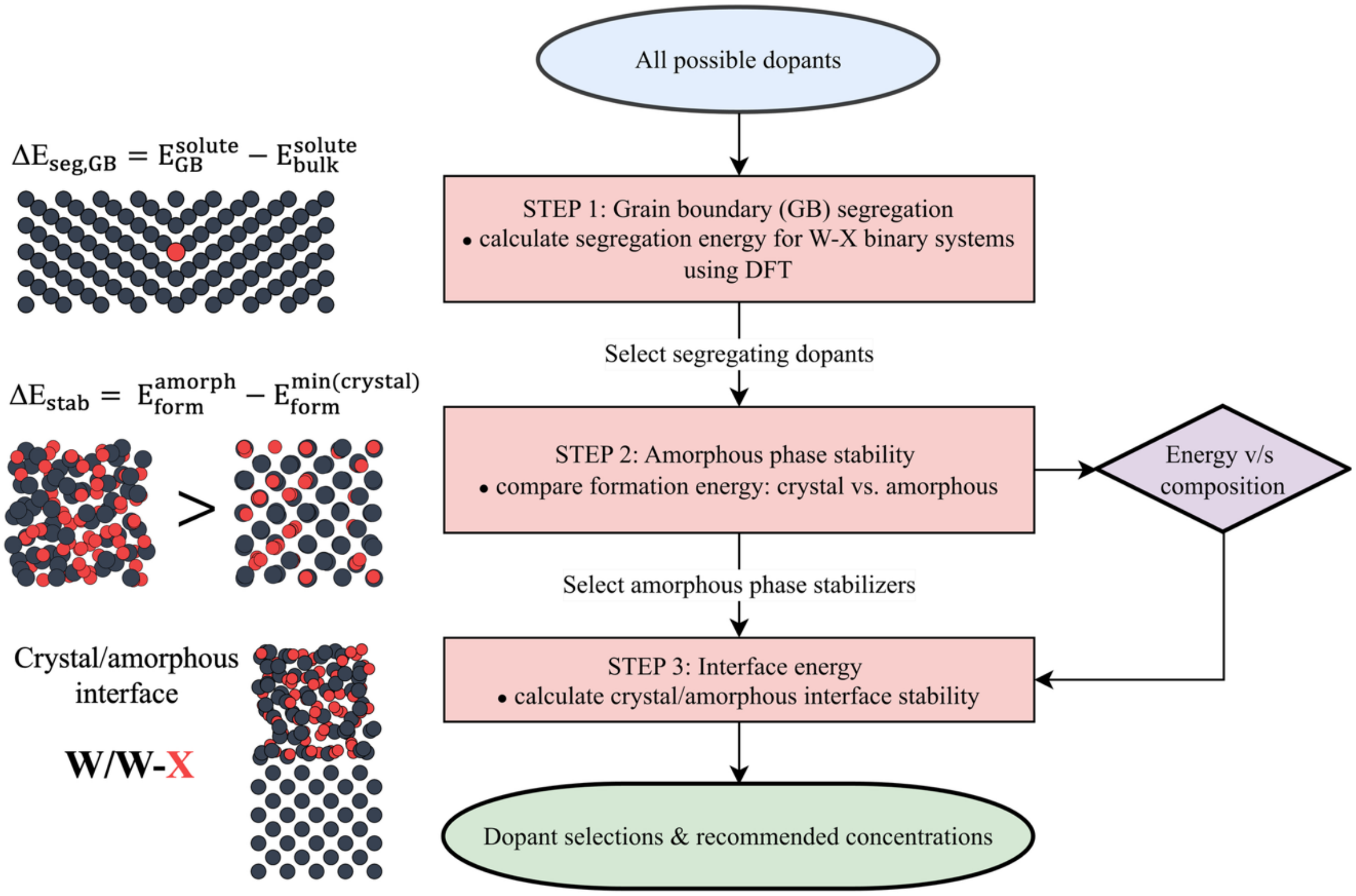

**Figure 1: Schematic illustration of the computational alloy design framework for predicting amorphous grain boundary complexions. The framework employs first-principles calculations to screen segregating dopant, evaluate phase stability through formation energy differences between crystalline and amorphous structures, and compute interface energetics to identify promising alloy candidates for experimental validation.**

All first-principles calculations were performed using the Vienna Ab initio Simulation Package (VASP) with the projector augmented wave (PAW) method.[43,44] The generalized gradient approximation (GGA) as parameterized by Perdew, Burke, and Ernzerhof (PBE) was employed for the exchange-correlation functional.[45] A plane-wave cutoff energy of 520 eV was used throughout all calculations to ensure convergence. A smearing width of 0.1 eV with the Monkhorst-Pack scheme is utilized to create a k-point mesh for Brillouin zone integration, with a precision of 0.03.[46] The electronic self-consistency loop was converged to $10^{-6}$ eV, while ionic relaxations were performed until forces on all atoms were below 0.01 eV/Å. Spin-polarized calculations were employed for all systems containing magnetic elements (Fe, Co, and Ni) to accurately capture their electronic structure.

Scheiber et al.[47] systematically investigated the segregation behavior of 3d, 4d, and 5d transition-metal dopants at Σ3(111)[1-10] high angle grain boundaries in W using density functional theory (DFT) calculations. These authors computed segregation energies for three distinct sites at the grain boundary relative to a bulk site. The site with the lowest segregation energy was identified as the preferred segregation site for each dopant. Further, to calculate the amorphous stabilization energy, both an amorphous state and a reference crystalline state are needed. For the amorphous structures, initial configurations were generated by melting the target composition with ab initio molecular dynamics (AIMD) simulations. Systems containing 128 atoms were heated to 5000 K and equilibrated for 5 ps using NPT ensemble with a Langevin thermostat with a standard 1 fs timestep, followed by a subsequent relaxation at 0 K. The same energy and force convergence criteria described above were used for this relaxation step. This resulted in a quenched amorphous structure that is representative of the premelting environment. For the crystalline state, our primary focus was on body-centered cubic (BCC) states because this is the equilibrium structure of W. Despite being defects, grain boundaries typically retain structural features derived from the parent lattice, as has been shown for high-angle grain boundaries.[48,49] For heavily doped states, where the majority element is no longer W, it is conceivable that the competing crystalline phase would be either face-centered cubic (FCC) or hexagonal close-packed (HCP), so additional crystalline phase energies were calculated in those cases. In general, the comparison is between the amorphous phase and the lowest energy crystalline phase.

Crystal-amorphous interfaces were modeled using selective dynamics supercells containing approximately 256 atoms. The interface structures consisted of frozen BCC W crystal oriented on the (001) lattice plane in contact with an amorphous region of the target grain boundary concentration. The interfacial energy was calculated as:

$$\gamma = \frac{E_{\text{interface}} - E_{\text{crystal}} - E_{\text{amorphous}}}{2\text{A}} \quad (5)$$

where $E_{\text{interface}}$ is the total energy of the interface system, $E_{\text{crystal}}$ and $E_{\text{amorphous}}$ are the energies of the corresponding bulk phases, and $A$ is the interface area. The factor of 2 accounts for the two interfaces that are present in a periodic supercell. Equation (1) also requires the energy of a clean W grain boundary. Here, a Σ3(111)[1-10] boundary was simulated, serving as a representative high-angle grain boundary, with DFT and a grain boundary energy of 0.198 eV/Å$^2$ was calculated, which is in good agreement with previously reported literature values.[50] This grain boundary energy falls within the typical range of 1.8–2.8 J/m$^2$ exhibited by the majority of high-energy grain boundaries in W,[50] ensuring that the thermodynamic driving force for complexion formation captured in our simulations is representative of a substantial fraction of the grain boundary population in polycrystalline microstructures. Although all of our calculations are at 0 K, this approach is still justified because it provided a conservative set of predictions. Elevated temperatures generally promote disordering by increasing entropic contributions, meaning that if a system favors amorphization at 0 K then this tendency will only be enhanced at the processing temperatures where complexions form. To confirm such behavior, AIMD calculations were performed for one alloy system and dopant concentration level at 1600 K.

## 3. **Results and Discussion**

### 3.1. **Dopant Selection for Complexion Formation in W-Rich Binary Alloys**

A necessary condition for formation of amorphous grain boundary complexions is that dopant atoms first accumulate at the grain boundary. Negative segregation energies indicate thermodynamically favorable segregation, where dopants preferentially occupy grain boundary sites, while positive segregation energies would suggest dopant depletion at the boundary. This segregation energy also influences complexion stability through its effect on interfacial composition. Strongly segregating elements can achieve high grain boundary concentrations even at low bulk doping levels, potentially enabling complexion formation with minimal alloying. For the case of W, extensive first-principles calculations can already be found in the literature that have mapped the segregation energies of transition metal dopants at representative grain boundaries. If this data does not exist, DFT calculations of representative grain boundaries with dopant atoms at a grain boundary site can be compared to the same calculation with the dopant at a grain interior site, per Equation (3). Scheiber et al. [47] reveal that segregation energies are all negative yet span a wide range, with the strongest segregating spices having segregation energies below -2 eV, as shown in **Figure 2**. Several important trends emerge from this data. Group III, VIII, IX and X transition metals exhibit strong segregation, namely Y, Fe, Co, and Ni as well as the noble metals. Refractory elements show weaker or unfavorable segregation. This behavior reflects their electronic and atomic size similarity to W, which reduces the energetic penalty for bulk substitution. The elastic contribution to segregation energy, arising from dopant-host size differences, plays a critical role in this behavior.[51-53] Grain boundaries provide sites with varying local atomic volumes, allowing oversized or undersized dopants to reduce elastic strain energy through segregation at specific sites.[51] The trend reveals that the classical activated sintering aids for W, specifically Ni and Co, are predicted to exhibit strong grain boundary segregation.[10,24,29]

However, grain boundary segregation alone is insufficient to fully explain this behavior, as elements such as Ir [54] and Cu [10,24] also exhibit segregation but do not enhance the sintering of W.

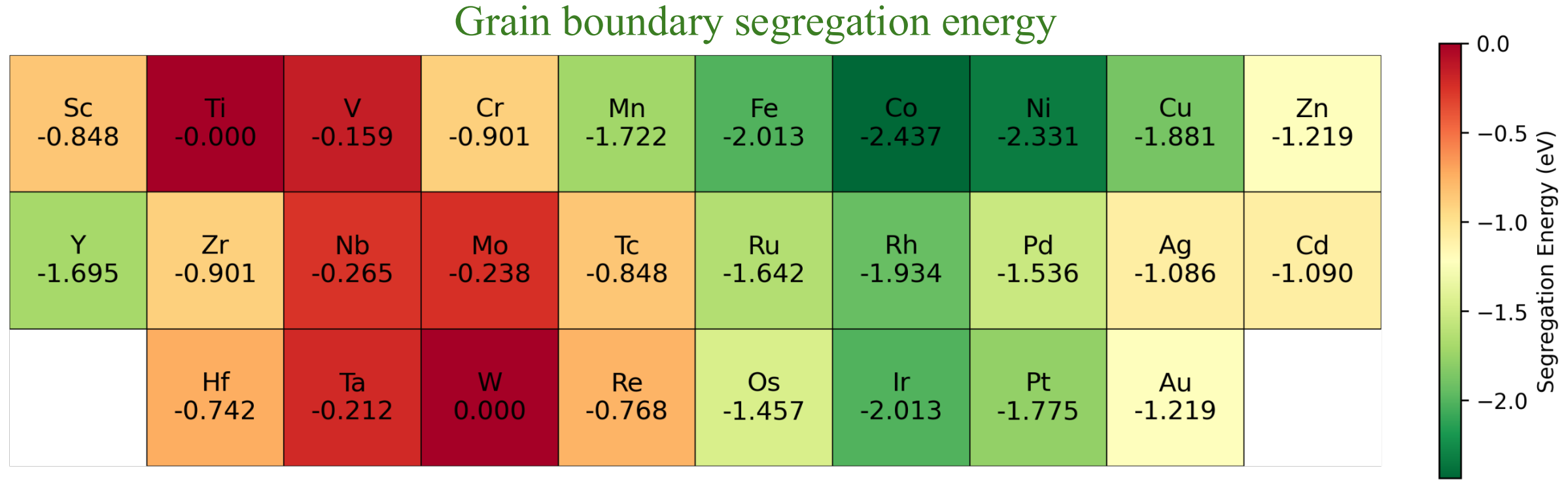

**Figure 2: Heatmap of grain boundary segregation energies for all transition metal dopants in W, with data extracted from Ref.[47].**

An element must both segregate to the grain boundary and promote structural disorder to enable amorphous complexion formation via premelting. The amorphous and crystalline structures used to assess stability are shown in **Figure 3(a)**. An equiatomic W–50 at. % X composition is first used to represent the highly enriched local chemistry expected once strong segregation has occurred, with this assumption later relaxed to explore various grain boundary concentrations. Equiatomic or near-equiatomic segregation levels at the grain boundary have been observed in a range of metal systems, including W-, Pt-, and Cu-rich nanocrystalline alloys [55-58]. An equiatomic composition is also a reasonable starting point because it serves as a baseline for solid-state amorphization via mechanical alloying.[59] An equiatomic composition maximizes configurational entropy, which is believed to stabilizes crystalline solid solutions, and simultaneously allows us to evaluate the threshold at which atomic size mismatch and structural distortion overcome the entropy-driven stabilization effect, ultimately driving the ordered crystal into an amorphous state.[60,61] This enables a direct comparison between ordered and amorphous structures without explicitly modeling all varieties of grain boundaries, allowing for rapid computation. The formation energy difference between amorphous and crystalline phases for W-doped with transition elements from the 3d, 4d, and 5d series is shown in **Figure 3(b)**. The calculations reveal large variations across the periodic table, with some large positive amorphous stabilization energies, suggesting a highly unstable amorphous phase, while other dopants promise low or even negative amorphous stabilization energies, signaling promising alloy combinations.

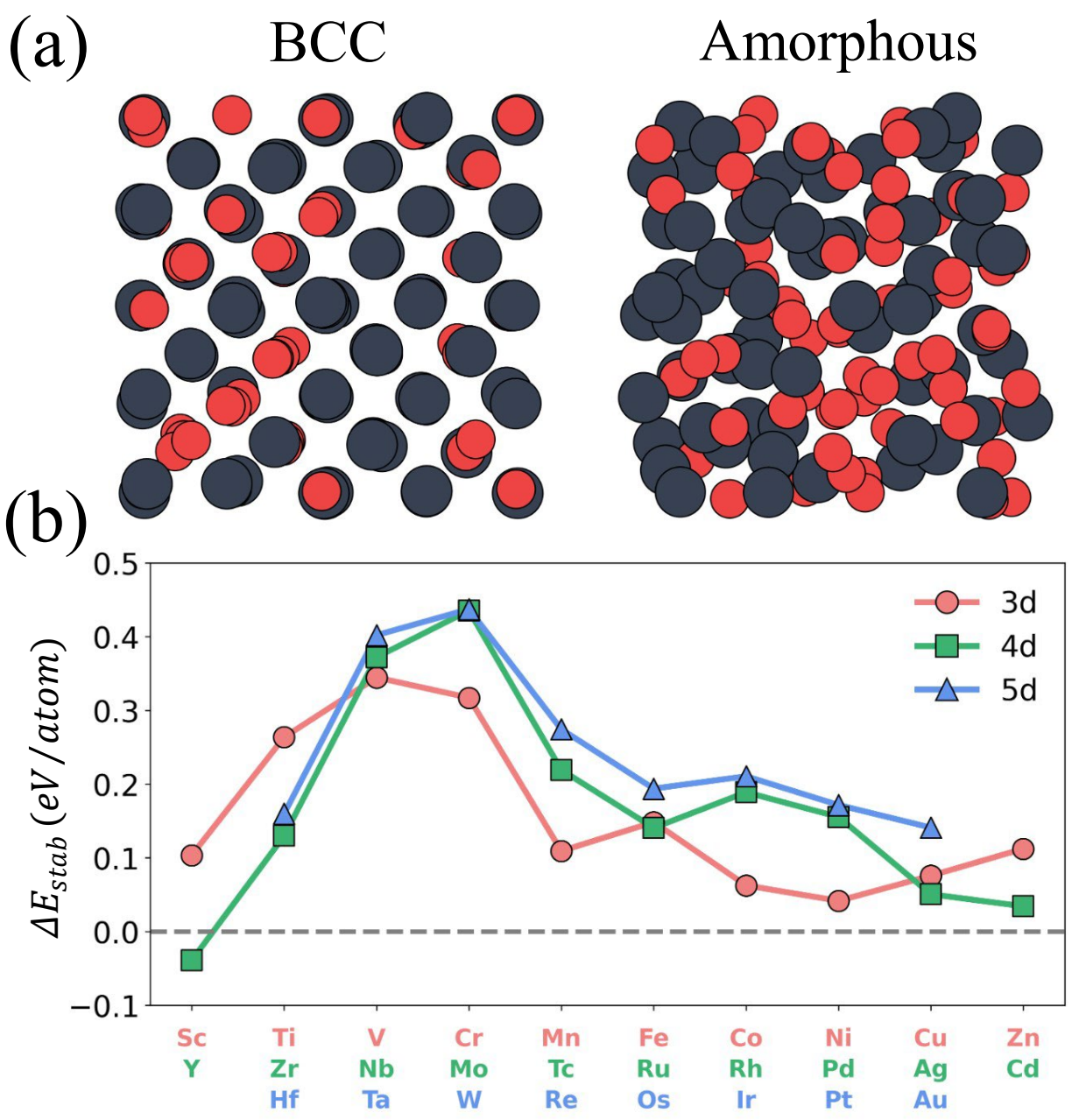

**Figure 3: (a) Structures of crystalline and amorphous phases of doped W that were used for DFT calculations. Gray atoms represent W and red atoms represent dopants (Ni, in this example). (b) Energy difference between the amorphous and the most stable crystalline phase for W doped with 3d, 4d, and 5d transition elements.**

**Figure 4** presents the amorphous stabilization data in a periodic table format. Among all of the transition metals, Y yields the lowest energy difference (dark green) and is in fact negative, indicating a natural preference for the amorphous structure. In contrast, Co, Ni, Cu, Ag, and Cd exhibit energy differences that are positive yet close to zero (green), suggesting near-degenerate stability between the crystalline BCC and amorphous states. Consequently, a modest external energy input, such as that provided by mechanical alloying[42] or from the replacement of the grain boundary with amorphous-crystalline interfaces may overcome this small energy barrier and stabilize the amorphous phase/complexion. Among this group, Y, Co, and Ni exhibit strong segregation. On the other hand, Ag and Cd possess favorable amorphous stabilization energies yet do not segregate strongly to W grain boundaries, meaning they may not accumulate at sufficient concentrations to enable amorphous complexion formation. Similarly, while Cu exhibits favorable segregation behavior, it has been shown experimentally to have no effect on the sintering kinetics of W[24], indicating that segregation and low amorphous stabilization energy together are still not sufficient in this case. An additional consideration for effective activated sintering is that the processing temperature must be high enough for appreciable diffusion in both the dopant-rich grain boundary region and the W matrix. The low melting point of Cu (1358 K) would suggest that amorphous complexions should form at ~800-860 °C, but W exhibits negligible atomic mobility (~22% of the melting point of W). In contrast, effective dopants such as Ni, Co, and Y could form amorphous complexions at ~1200 °C or ~1473 K (~40% of the melting point of W), where W diffusion becomes appreciable. As activated sintering is a secondary signal of the complexion formation, it is indeed important to keep these extra kinetic considerations in mind. Fe is included as a moderately favorable dopant because it combines strong grain boundary segregation with a reasonably low amorphous stabilization energy; although its energy difference is larger than the most favorable dopants (Y, Co, Ni) it sits at a crucial spot between dopants that promote versus inhibit complexion formation.

In contrast, refractory dopant species like Mo, Ta, and Nb (dark red) possess high positive values that suggest an unstable amorphous phase or a highly stable BCC phase. The individual trends for 3d, 4d and, 5d elements show higher values at $d^{3,4}$ configuration states which are similar to that of W($5d^4 6s^2$) and leads to a naturally tendency for strong bonding and ordering. These elements exhibit strong *d*-band bonding and good electronic/size compatibility with W, which stabilizes the parent BCC phase and makes amorphous phase formation energetically unfavorable.[62-64] Conversely, early transition metals (with largely empty *d*-bands) and late transition metals (with nearly filled *d*-bands) exhibit electronic configurations that are less compatible with the strong directional *d*-band bonding characteristic of W.[62-64] This reduced hybridization, combined with significant atomic size differences, reduces the energetic preference for crystalline ordering and thus lowers the energy difference between amorphous and crystalline phases.[65]

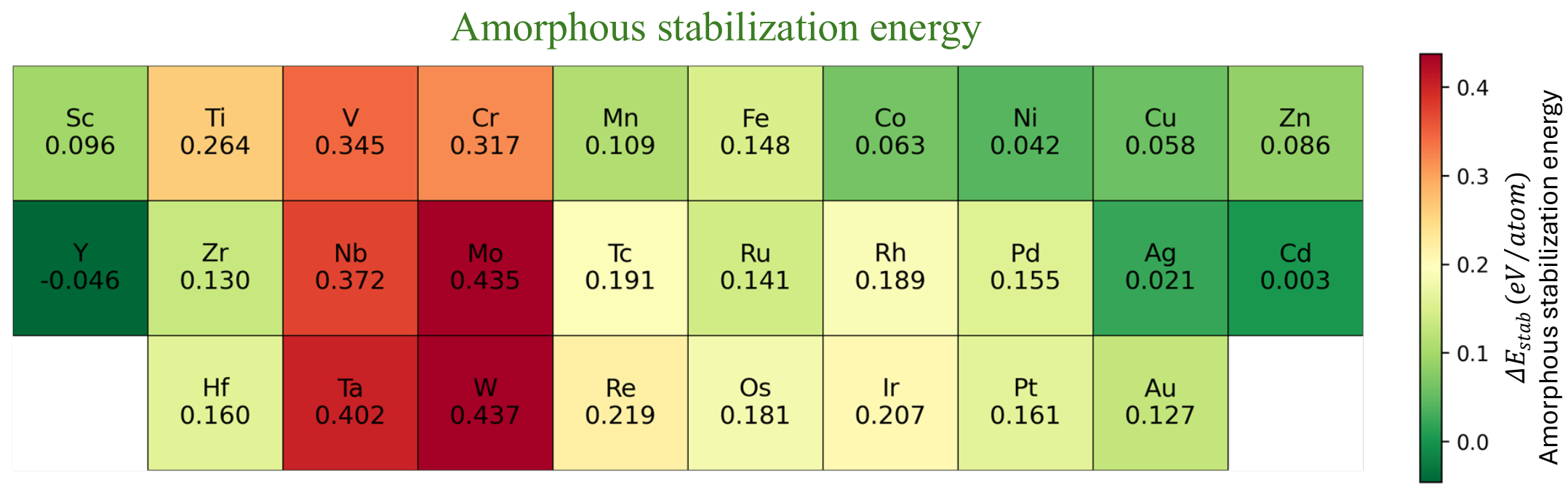

**Figure 4: Heatmap of calculated amorphous stabilization energies for all transition metal dopants. Lower values (green) indicate greater thermodynamic favorability for amorphous complexion formation. A selection of promising dopants (Y, Co, Ni, Fe) is shown in green, while poor candidates (Mo, Ta) are shown in red.**

The assumption of equiatomic concentration can next be relaxed to explore how grain boundary composition affects the stabilization of the amorphous phase. **Figure 5** presents the energy of amorphous and crystalline phases for selected dopants representing both favorable (Y, Co, Ni), marginally favorable (Fe) and unfavorable (Mo, Ta) dopant selections. In these figures, the energy of pure W in BCC structure is taken as the zero energy reference. For favorable dopants shown in **Figure 5(a-c)**, the formation energies for crystalline and amorphous structure are closest for concentrated compositions, typically between 40 and 70 at.% dopant. This behavior suggests the existence of an optimal dopant concentration range at grain boundaries where amorphous complexions are most likely to form. For the W–Y system in **Figure 5(a)**, the amorphous phase is even energetically preferred over the crystalline phase in this concentrated composition range. W-Fe in **Figure 6(d)** shows its smallest difference in the intermediate concentration range, yet the difference is 2-3 times than that of the favored dopants and hence is labeled in yellow. For the four binary alloys in **Figures 5(a-d)**, having too few dopants means that the amorphous phase is relatively unstable compared to the crystalline phase and amorphous complexion formation would be unlikely. Hence, grain boundary concentration needs to be relatively high and significant segregation must occur. Unfavorable dopants such as Mo and Ta in **Figures 5(e) and (f)** do not exhibit any composition range where the amorphous phase becomes energetically preferred. Instead, the BCC structure remains significantly more stable across all compositions.

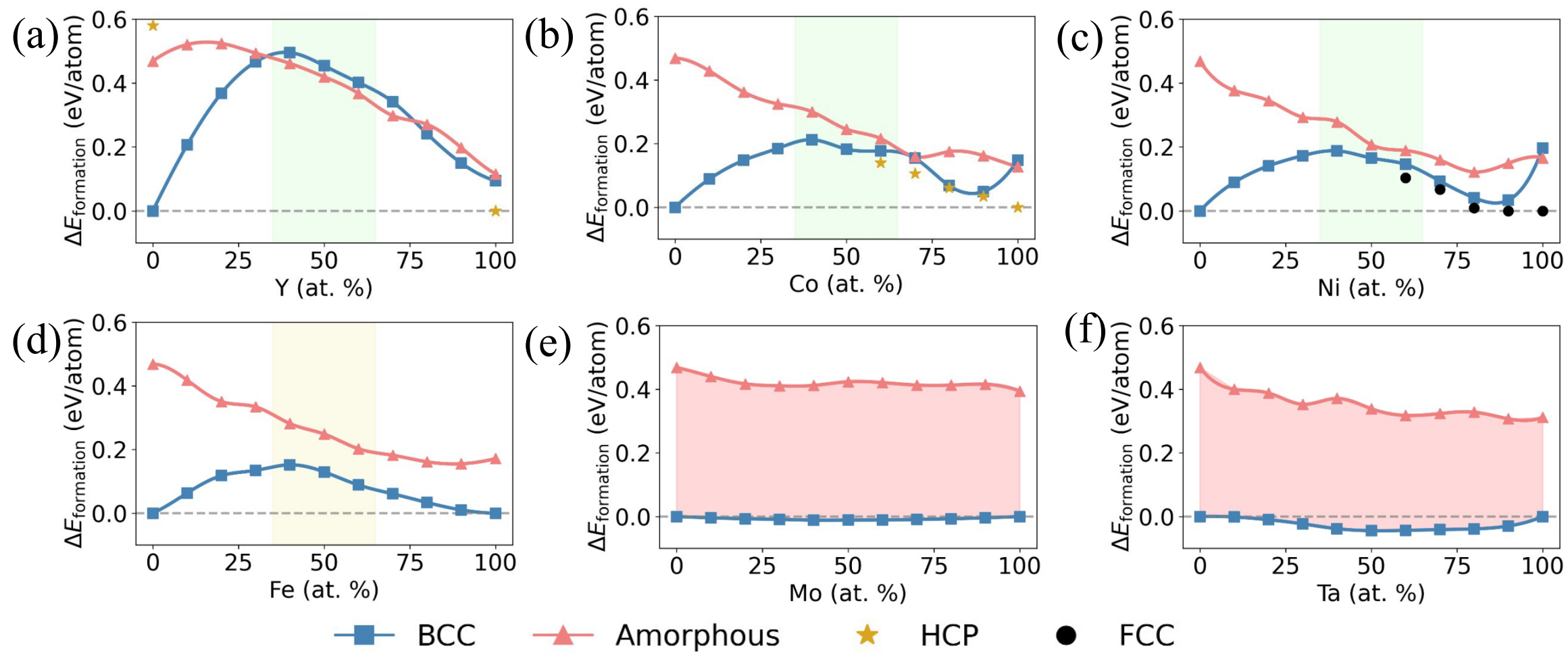

**Figure 5: Difference in formation energies ($\Delta E_{formation}$, with pure W taken as the reference state) between crystalline and amorphous phases as a function of dopant concentration for selected transition metals. (a-d) Favorable dopants (Y, Co, Ni) and moderately favorable dopants (Fe) exhibit the most stable amorphous phase at intermediate compositions (shaded in green and yellow, respectively). (e-f) In contrast, unfavorable dopants (Mo, Ta) do not stabilize the amorphous phase, with BCC remaining the lowest energy structure across all concentrations by a large margin (shaded in red).**

The data above tells one (1) which dopants will segregate and (2) which dopants are promising for hosting amorphous structures. However, most of the favorable alloys still had a positive amorphous stabilization energy, which must be overcome by a reduction in interfacial energy if an amorphous complexion is to be stable. Such a reduction in interfacial energies can potentially be achieved by replacing the original high-energy crystalline grain boundary with two lower-energy crystalline-liquid (crystalline-amorphous, after the sample is quenched) interfaces.[1,9,11,28] A representative crystal-liquid interface is shown in **Figure 6(a),** consisting of crystalline BCC W in contact with an amorphous W-X alloy. The most straightforward approach to evaluate interfacial stability would be to perform a standard structural relaxation allowing all atomic positions and cell dimensions to optimize simultaneously. However, such a direct method proved problematic for several systems, as rapid structural transformations during relaxation obscured the target amorphous-crystalline interface configuration. For example, the W/W-Mo interface underwent complete recrystallization, transforming the initially amorphous region into an ordered structure. This rapid ordering certainly indicates that amorphous complexions are unstable in the W-Mo system, yet precludes quantification of the targeted interfacial energies for analysis. In contrast, W/W-Ni and W/W-Y interfaces maintained their amorphous character after relaxation, though with different degrees of structural modification. The W/W-Y interface showed notable structural distortion at the crystal-amorphous boundary, suggesting accommodation of the lattice mismatch through local atomic rearrangements. To maintain a consistent comparison between different alloys, a constrained relaxation approach was adopted where selective dynamics fixed the atomic positions of the pure W crystal in two directions while allowing full relaxation in the direction perpendicular to the interface. This approach preserved the amorphous-crystalline character of the interface while allowing the system to reach a representative local energy minimum. The calculated interfacial energies are compiled in **Figure 6(b).** The favored alloy systems from **Figure 5** (W-Co, W-Ni, and W-Y) have low values (green), indicating promise for amorphous complexion formation because the crystal-liquid interfaces have relatively low energies. In contrast, large positive values for W-Mo and W-Ta reinforce that these dopants would be less likely to experience amorphous complexion formation since their crystal-liquid interfaces also have relatively high energies.

Although, dopants such as Y, Ni, Co, and Fe reduce the crystal-amorphous interface energy below the clean grain boundary baseline of 0.198 eV/Å², it is important to note that complexion formation replaces a single grain boundary with two crystal-amorphous interfaces (Equation 1). This geometric factor of two requires that the interface energy must be reduced to less than half the clean grain boundary energy for a net interfacial energy benefit, although this is at elevated temperature (as discussed in the Methods section). **Figure 6** can still be used to understand the relative importance of the amorphous-crystalline interface energies and to rank perspective alloys. Here, the amorphous-crystalline interface energy varies roughly ~0.05 eV/Å² across all dopants (from 0.151 eV/Å² for Y to 0.201 eV/Å² for W), remaining close to the clean grain boundary energy of 0.198 eV/Å². The amorphous stabilization energy therefore is the dominating factor as it varies over a much larger range, going from -0.046 eV/atom for Y to 0.437 eV/atom for W. As for dopant selection, one should prioritize elements with low amorphous stabilization energies and also low amorphous-crystalline interface energies. The most favorable dopants (Y, Co, Ni) all exhibit amorphous stabilization energies below 0.07 eV/atom, while unfavorable dopants (Mo, Ta) exceed 0.4 eV/atom. The favorable dopants also have relatively low amorphous-crystalline interface energies compared to the unfavorable dopants. We note that while these metrics all point in the same direction for W-rich alloys, these calculated energies can have different importances and trends for other alloy systems. We note that the (001) orientation was used for the BCC W crystal-amorphous interface; other crystallographic planes may yield different interfacial energies.

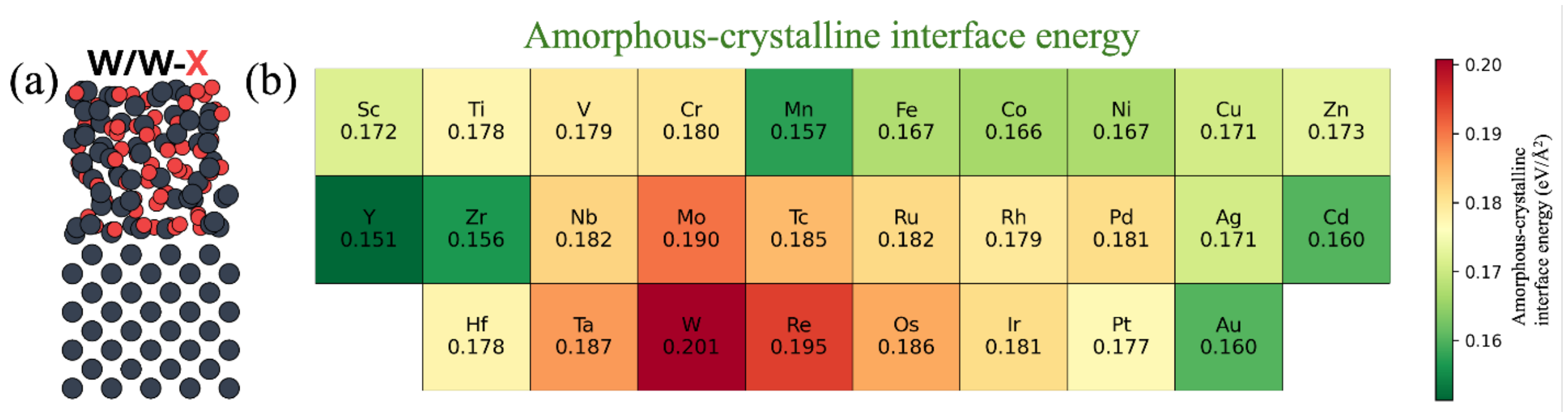

**Figure 6: Interfacial energy determination of a BCC W and amorphous W-X alloy interface. (a) Initial configuration showing a heterostructure of crystalline BCC W (bottom) and amorphous W-X alloy (top), where X represents different alloying elements. The grey atoms are W and red are Ni in this specific example. (b) Heat map displaying calculated interfacial energies (eV/Å²) for various W/W-X combinations, with lower values (green) indicating favorable interface formation and higher values (red) indicating unfavorable interfaces.**

The discussion above demonstrates that Y is expected to be the best dopant choice for amorphous grain boundary complexion formation in a binary W-rich alloy. If the DFT calculations performed so far are plugged into Equation (1), a complexion thickness of -0.9 nm would be predicted for W-Y. While this suggests complexions would not form, one must recall that these 0 K calculations do not account for the effect of temperature. To achieve a more accurate value, we calculated values of γ and $\Delta E_{stab}$ for W-Y at 1600 K using AIMD and the complexion thickness was found to increase to +1.31 nm for an equiatomic grain boundary concentration. AIMD calculations at 1600 K were also performed for W-Ni and W-Co, giving positive complexion thicknesses of 0.21 nm and 0.23 nm, respectively. The exact predictions for all three alloy systems are shown in **Figure 7**. Importantly, the relative ranking of dopants is preserved. Y produces the thickest complexions, consistent with its lowest $\Delta E_{stab}$, followed by Co and Ni. This demonstrates that 0 K calculations, while not yielding physically meaningful absolute thicknesses, are able to correctly identify favorable dopants based on their relative $\Delta E_{stab}$ values. The less computationally expensive 0 K calculations remain very useful for initial alloy screening, which can help users rapidly shortlist dopants. Once dopants are selected, computationally expensive AIMD can give final predictions. At 1600 K, the amorphous-crystalline interface becomes more stable and the interfacial penalty is reduced by ~25% compared to the value at 0 K. We note that the exact value of the predicted complexion thickness will depend on the energy of the grain boundary in question. Here we use a boundary that is representative of high-angle and high-energy boundaries in W. However, low-energy boundaries such as coherent twin boundaries would be much less likely to transition to an amorphous structure according to Equation 1. Therefore, the fraction of boundaries that transform to amorphous complexions will depend on the exact grain boundary concentration of dopants, temperature, and individual boundary energies.

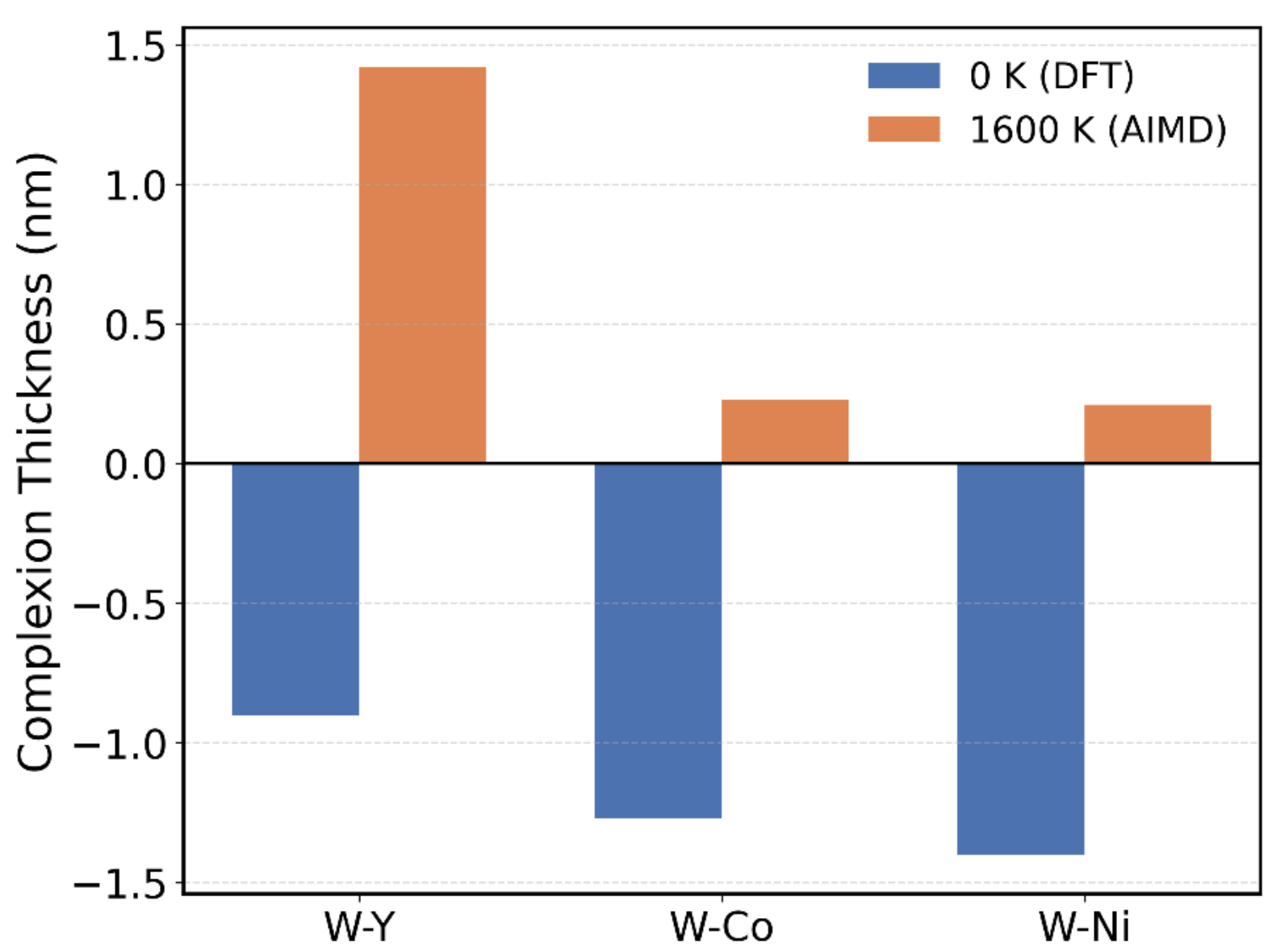

**Figure 7: Calculated complexion thickness at 0 K and 1600 K for W-Y, W-Co, and W-Ni alloys with equiatomic grain boundary concentrations. Negative thicknesses at 0 K become positive at 1600 K, confirming the prediction of stable amorphous complexion formation at elevated temperatures.**

### 3.2. Stabilization Mechanisms for Amorphous Complexion Structures

To explore mechanistic explanations for why some dopants are good at stabilizing the amorphous phase, which again appears to be the variable that is most strongly variable for W-rich alloy selection, one can investigate atomic packing and the effect of doping on the lattice. **Figure 8** presents equilibrium configurations for W-Mo and W-Y crystalline states, providing representative examples of systems which are favorable and unfavorable for amorphous complexion formation. The W-Mo system (**Figure 8(a)**) maintains an ideal BCC structure with atoms occupying their expected lattice sites, demonstrating excellent structural compatibility between W and Mo. In contrast, the W-Y system (**Figure 8(b)**) displays significant local lattice distortions, with atoms substantially displaced from ideal BCC lattice positions due to the large atomic size mismatch between the atomic species. These distortions lower the stability of the crystalline phase relative to the amorphous state, thereby promoting complexion formation.

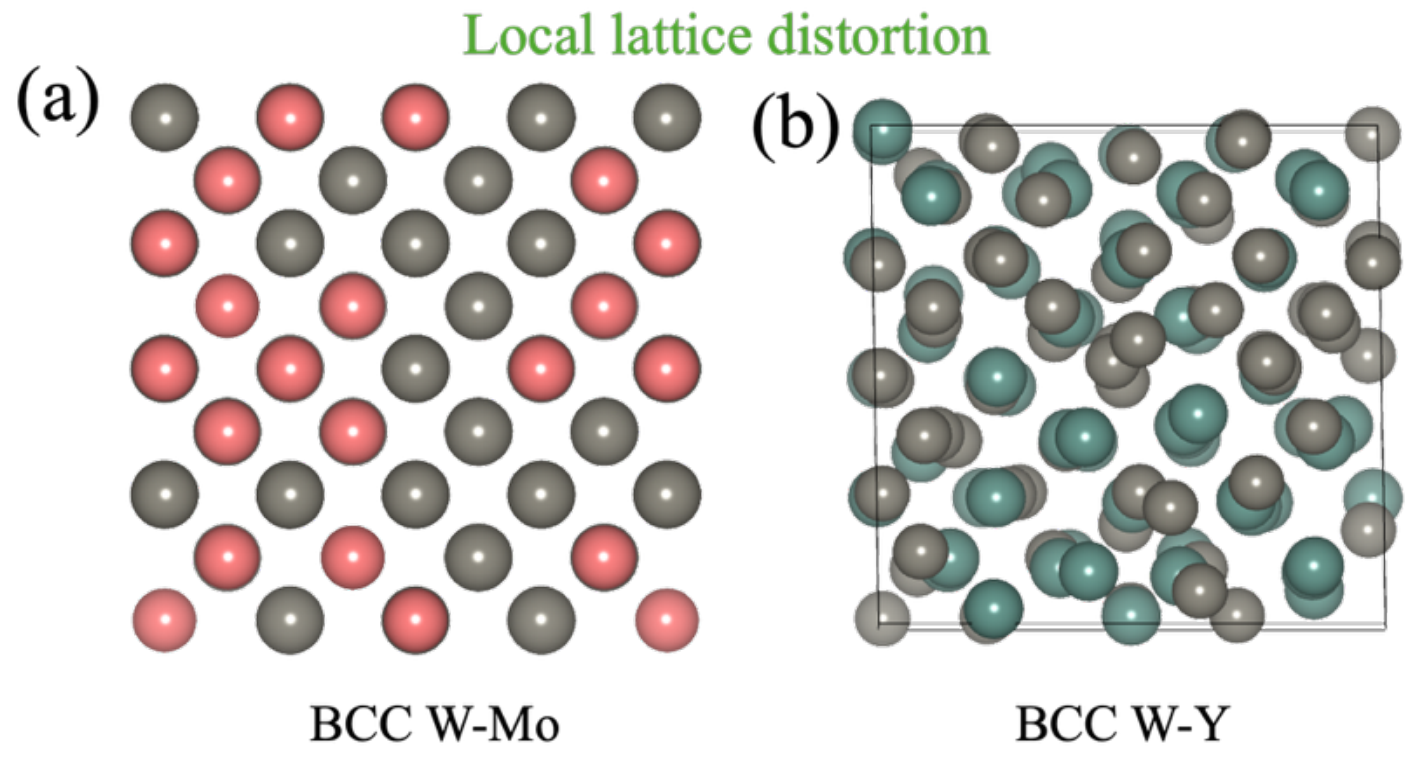

**Figure 8: Local lattice distortion in BCC W-rich alloys. (a) BCC W-Mo maintains an ideal BCC structure with minimal lattice distortion. (b) BCC W-Y displays significant local lattice distortions, with Y atoms substantially displaced from ideal lattice positions, suggesting a propensity for amorphization. The contrasting behaviors provides a mechanistic explanation for why W-Mo forms stable crystalline structures while W-Y tends toward structural disorder.**

Electronic structure and its distribution can also provide insight into amorphous stabilization relative to the crystalline state. **Figure 9** compares the electronic density of states (DOS) profiles of the amorphous and BCC structures for the alloys identified in **Figure 5**. For favorable dopants in **Figures 9(a-c)**, the DOS of amorphous and crystalline phases are very similar, with nearly identical distributions when comparing within a single atomic species (e.g., W in amorphous state versus W in the crystalline state). The W DOS curves exhibit a characteristic pseudogaps near the Fermi level in all systems except for BCC W-Mo and W-Ta, where the alloys are extremely stable in BCC structure. For favorable dopants, the dopant DOS shows only subtle differences between crystalline and amorphous phases. In the W-Co system in **Figure 9(b)**, the Co DOS displays a well-defined peak between -2 to -1 eV in the BCC structure that gets smeared out in the amorphous structure. The W-Ni system in **Figure 9(c)** exhibits similar behavior between -3 to -1 eV. The W-Y system shown in **Figure 9(a)** shows minimal change between the BCC and amorphous structures. The W-Fe system in **Figure 9(d)** presents an intermediate case, where the sharp, well-defined Fe peaks between -4 to -1 eV in the BCC structure broaden more in the amorphous structure, as compared to Co and Ni. In contrast, for unfavorable dopants (Mo, Ta) shown in **Figures 9(e-f)**, the BCC structure shows several sharp, well-defined peaks, a characteristic of stable ordered structure, while the amorphous structure shows significant broadening and smearing of these features. In these alloys, the BCC structure enables favorable electronic interactions and "resonance" within the *d*-band that cannot be maintained in a disordered amorphous state, thus incurring a high energetic cost for amorphization.

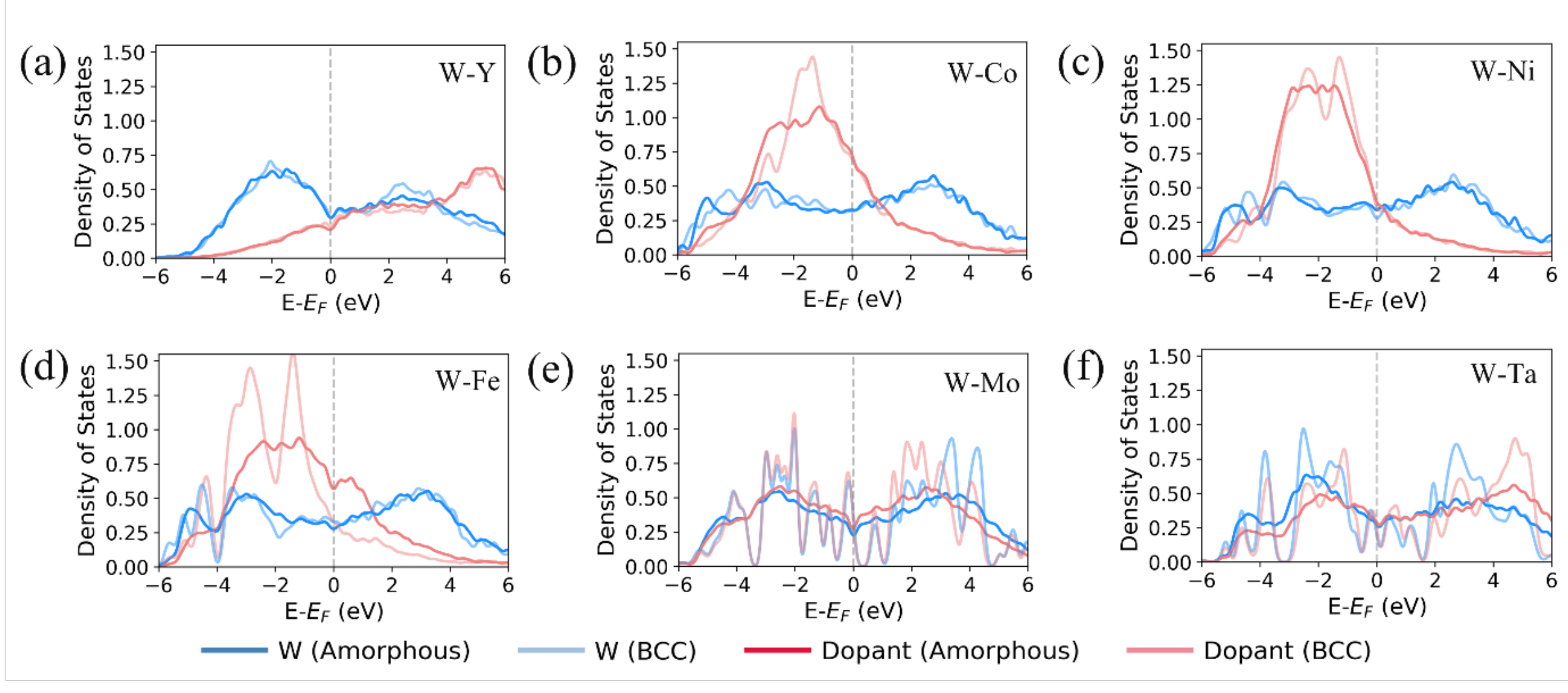

**Figure 9: Electronic density of states for binary W systems in the amorphous and BCC phases. (a-d) For favorable dopants (Y, Co, Ni, Fe), the DOS profiles of the amorphous and BCC phases are notably similar near the Fermi level ($E_F$), indicating comparable electronic stability. (e-f) In contrast, for unfavorable dopants (Mo, Ta), the DOS profiles of amorphous and BCC phases diverge significantly, with the BCC phase exhibiting lower states at $E_F$, suggesting superior thermodynamic stability.**

### 3.3. Benchmarking and Extension to More Chemically Complex Alloys

It is important to benchmark the predictions above against experimental data from the literature. Direct identification of amorphous grain boundary complexions exists for W-Ni, where ~0.6 nm thick complexions were observed with high resolution transmission electron microscopy.[11,30] While examples of such direct visualization do not exist for all of the alloys of interest, evidence can be found from sintering studies as amorphous complexions have been put forth as the mechanism for activated sintering. Hayden and Brophy[23] first reported activated sintering in W, while Luo et. al.[30] showed that this enhancement occurs because dopants segregate to grain boundaries to cause boundary premelting below the bulk eutectic temperature. The formation of such amorphous complexions provides rapid mass transport pathways, enabling densification at temperatures hundreds of degrees below those required for pure W. **Figure 10** shows the sintering onset temperatures ($T_{onset}^{sintering}$) of a number of W-rich alloys and pure W as a function of the amorphous stabilization energy. Strikingly, favorable dopants such as Co, Ni, and Y have the lowest $\Delta E_{stab}$ values and correspondingly lower $T_{onset}^{sintering}$.[10,24,29,66,67] The reduction in $T_{onset}^{sintering}$ is up to 1000 °C for the most effective dopants Ni and Y. Importantly, elements such as Mo and Ta that have high positive $\Delta E_{stab}$ values and are predicted by our framework to be poor candidates cluster in the upper right region of **Figure 10**, as only very minor reductions in sintering onset temperature are reported compared to pure W. Thus, there is a clear correlation between our calculated complexion stability and experimental sintering behavior validates that $\Delta E_{stab}$ is a dominant factor for grain boundary complexions and captures the relevant interfacial thermodynamics.

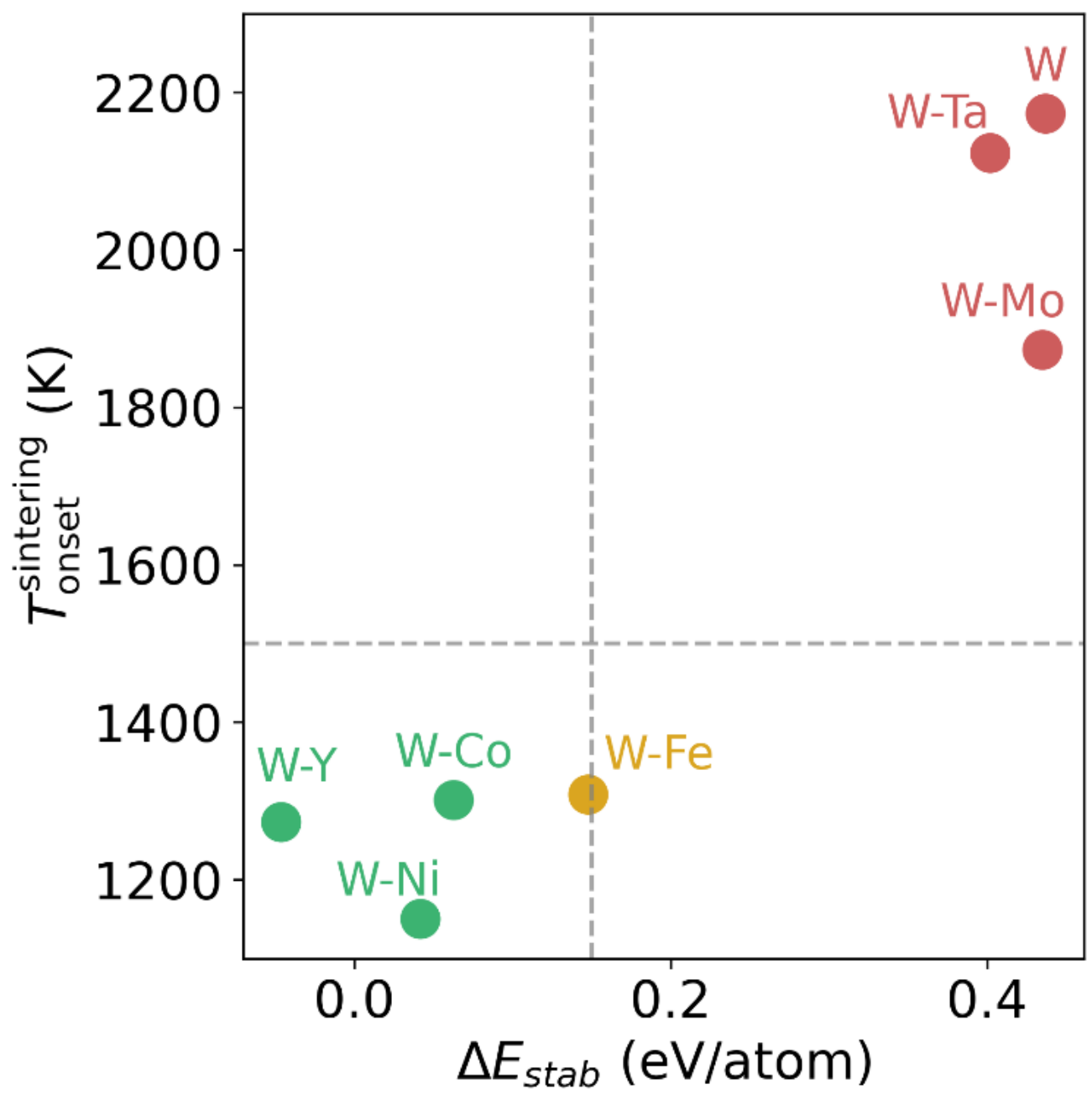

**Figure 10: Correlation between calculated amorphous stabilization energy ($\Delta E_{stab}$) and sintering onset temperature ($T_{onset}^{sintering}$) showing that alloys with lower amorphization stability (higher $\Delta E_{stab}$) require higher temperatures to initiate sintering. Dopant levels were ~0.6 at.% for Fe, Co, and Ni.[10,23,67] For comparison, Mo was added at ~9 at.%[68], Y at 1–10 at.%[67], and Ta at ~6 at.%.[69]**

The present alloy design framework based on amorphous stabilization energies provides an approach that is easily translatable to other materials, such as complex concentrated alloys where the base material is more chemically complex and where trial-and-error or empirical approaches are more common. In fact, first-principles calculations have been used to predict bulk amorphization in high-entropy alloys, where the amorphous phase was found to have the lowest formation energy relative to competing crystalline structures.[42] Concerning amorphous grain boundary complexions, recent work by Shivakumar et al.[31] demonstrated Ni-activated sintering in MoNbTaW. These authors proposed that amorphous complexions were present during processing to enable this behavior. In addition, Cao et al.[70] showed Co-induced activated sintering in equiatomic MoW, MoWNb, and MoWNbTa alloys at 1200 and 1300 °C. In these cases, Ni and Co were chosen based on prior observations of activated sintering of W-(Ni/Co)[24,30] and Mo-Ni[71] with minor doping. **Figure 11** shows BCC and amorphous simulated structures as well as the corresponding DOS for MoNbTaW and a Ni-doped MoNbTaW. The $\Delta E_{stab}$ for Ni-doped MoNbTaW was found to be 0.003 eV/atom, which is a significant reduction from the case of undoped MoNbTaW at 0.3725 eV/atom. These values align well with trends established in our binary W-X systems, with the Ni-doped MoNbTaW value falling in between W-Ni (slightly positive) and W-Y (slightly negative). As such, our computational framework would also predict the formation of amorphous complexions, matching observations in this more chemically complex alloy. The transferability of the amorphous stabilization energy to multicomponent systems suggests that the insights gained from simpler binary systems can guide dopant selection for engineering grain boundary complexions in compositionally complex alloys. Lattice distortions and electronic DOS also provide supporting mechanistic evidence to explain the formation of amorphous complexions in this system. **Figure 11(a)** shows almost no lattice distortion and well defined and sharp peaks are seen in the DOS in **Figure 11(b)** for the BCC MoNbTaW alloy, similar to trends in **Figures 9(e-f)** for binary W-Mo/Ta. The DOS curve is significantly different for the amorphous MoNbTaW in **Figure 11(d)**, suggesting a significant barrier for this disordering transition. In contrast, the lattice of the Ni-doped MoNbTaW is heavily distorted (**Figure 11(e)**) and the DOS for the BCC (**Figure 11(f)**) is very similar to the amorphous state (**Figure 11(h)**), resembling trends in the DOS curves for binary W-Ni in **Figure 9(c)**.

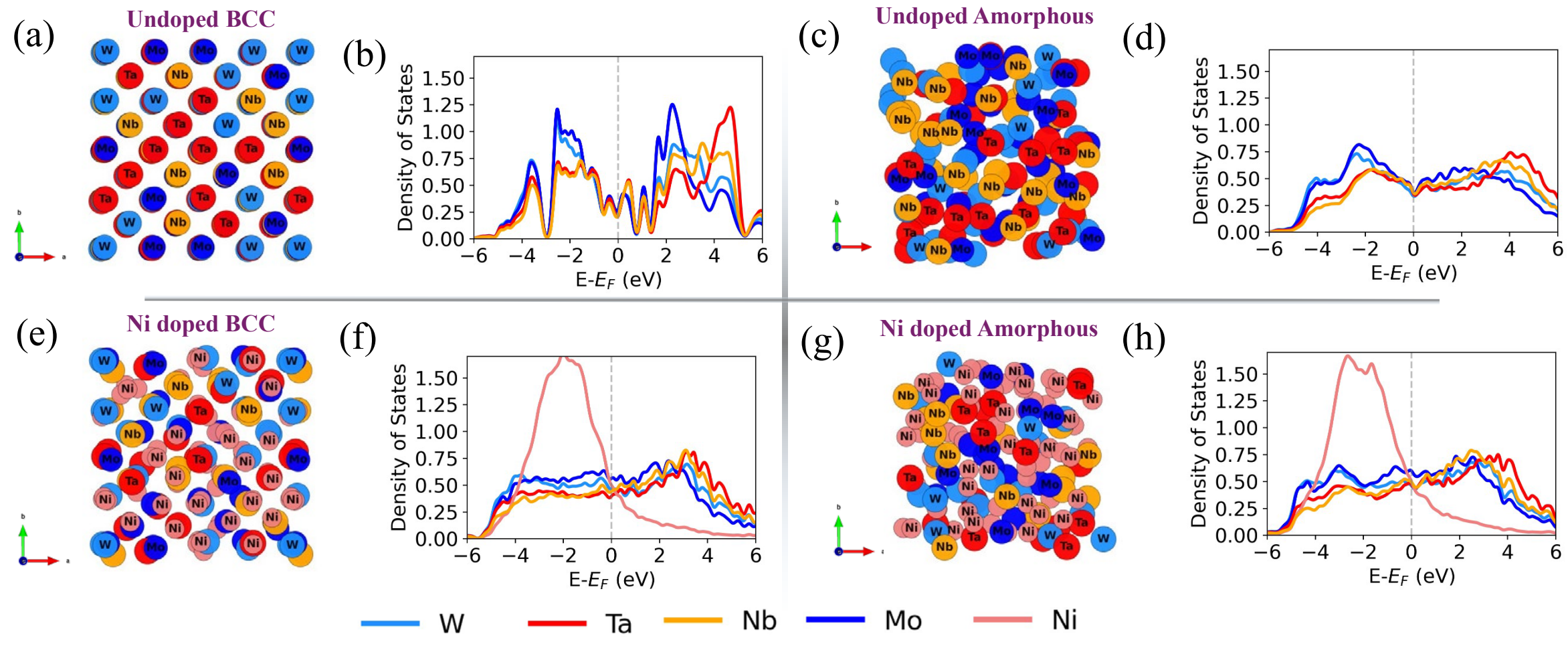

**Figure 11: Extension of our framework to Ni-activated grain boundary amorphization in MoNbTaW complex concentrated alloy. Atomic structures and density of states for (a,b) BCC MoNbTaW, (c,d) amorphous MoNbTaW, (e,f) Ni-doped BCC MoNbTaW, and (g,h) Ni-doped amorphous MoNbTaW. Significant lattice distortions and DOS similarity between crystalline and amorphous states in the doped alloy align with our earlier observations for binary alloys.**

Increased chemical complexity of the grain boundary region can also be probed with the framework developed here. Chemical complexity is an important concept for stabilizing metallic glasses, where increasing the number of constituent elements and their concentration generally enhances amorphous phase stability through topological frustration that inhibits crystallization.[72,73] To demonstrate the use of our framework on a multicomponent system, two ternary systems were chosen: (1) W-Y-Ni combines two dopants that individually promote amorphization, while (2) W-Y-Zr pairs an effective dopant (Y) with one that is unfavorable for complexion formation (Zr). Grain boundary compositions of 25-57 at.% dopant were explored to probe relatively concentrated compositions, although we note again that the bulk alloy compositions required to achieve these would be one to two orders of magnitude lower due to dopant enrichment to the interfaces. This capability is particularly valuable given that bulk phase diagrams become inadequate for designing activated sintering protocols in complex alloys. The W-Y-Ni system in **Figure 12(a)** shows generally low and often negative values of $\Delta E_{stab}$, suggesting a wide composition range where amorphous complexions can be stabilized. Unsurprisingly, Y is the dominant contributor, as compositions along the W-Y edge have lower stabilization energies than those along the W-Ni edge. However, the lowest $\Delta E_{stab}$ values do appear for the ternary compositions, with more Y than Ni, suggesting improvement with increased compositional complexity. In contrast, the data for W-Y-Zr alloys shown in **Figure 12(b)** has predominantly positive $\Delta E_{stab}$ across most of the ternary compositions. Here, the poor performance of Zr persists even when combined with effective Y. This confirms that simply adding more elements does not guarantee improved amorphization; ternary and higher combinations should be modeled in detail to understand the effectiveness for amorphous complexion stabilization.

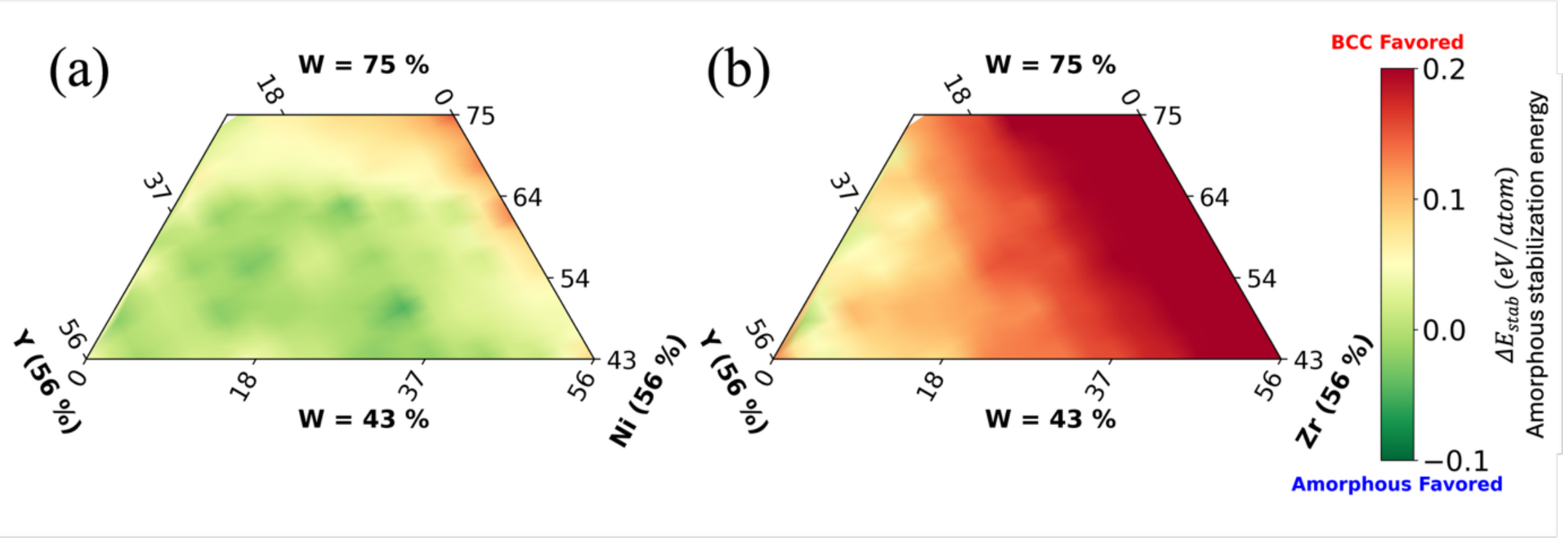

**Figure 12: Extension of $\Delta E_{stab}$ maps from binary to ternary alloys. Concentrated grain boundary compositions are shown for (a) W-Y-Ni and (b) W-Y-Zr alloys. Green regions (negative $\Delta E_{stab}$) favor amorphous complexion formation; red regions (large positive $\Delta E_{stab}$) favor crystalline structures.**

## 4. **Conclusions**

This work establishes a first-principles simulation framework for dopant selection and identification of ideal grain boundary concentrations to encourage amorphous complexion formation at grain boundaries. The framework evaluates multiple energy contributions (grain boundary segregation energy, amorphous stabilization energy, and amorphous-crystalline interface energy) to identify promising dopants. For the W-rich alloys used here as a representative example, the amorphous stabilization energy emerges as the dominant factor controlling dopant effectiveness, varying from -0.046 eV/atom for Y to 0.437 eV/atom for pure W. In contrast, the amorphous-crystalline interface energies remain relatively constant (~0.05 eV/Å² variation across all dopants) suggesting that dopant choice is less impactful here. The energy difference between amorphous and crystalline structures provides an easily computed metric that enables rapid screening decisions here, where low or negative $\Delta E_{stab}$ values identify dopants that promote grain boundary amorphization. For W, the most favorable dopants (Y, Co, and Ni) all exhibit amorphous stabilization energies below 0.07 eV/atom, while unfavorable dopants such as Mo and Ta are much higher and exceed 0.4 eV/atom.

Benchmarking against experimental activated sintering data validates the predictive capability of this framework, as the calculated $\Delta E_{stab}$ values correlate strongly with sintering onset temperatures for densification. The framework is easily extended to multicomponent systems where both the bulk and/or grain boundary regions have multiple atomic species. For example, Ni-activated sintering in MoNbTaW can be explained by the fact that $\Delta E_{stab}$ decreases from 0.3725 eV/atom for pure W to 0.003 eV/atom upon Ni doping. In general, the framework described here provides a rational, computationally efficient screening tool for identifying dopants that promote amorphous complexion formation in a wide variety of alloys, offering an alternative to trial-and-error experimental approaches. Future studies may focus on the performance of these amorphous grain boundary complexions, such as their ability to capture point defects created during irradiation. Here, the binding energies of point defects with the competing crystalline and amorphous structures would be of key importance and could enable the creation of models that can capture the long-term evolution microstructures containing amorphous grain boundary complexions.

**Acknowledgements**

This work was supported by the Advanced Research Projects Agency – Energy (ARPA-E) under the CHADWICK Program Award Number DE-AR0001993.